\documentclass[aip,apl,preprint,amsmath,amssymb,floatfix]{revtex4-1}
\usepackage{graphics}
\usepackage{graphicx}
\usepackage{dcolumn}

\usepackage[utf8]{inputenc}
\usepackage[T1]{fontenc}
\usepackage{mathptmx}
\usepackage{gensymb}
\usepackage[pdfusetitle,colorlinks=true,plainpages=false]{hyperref}
\usepackage{textcomp}

\hypersetup{citecolor=blue}
\hypersetup{linkcolor=blue}

\draft 

\begin{document}

\title{Temperature-Dependent Resistivity of Alternative Metal Thin Films}

\author{Marco Siniscalchi}
\altaffiliation{Present address: Department of Materials, University of Oxford, Oxford, United Kingdom}
\affiliation{Imec, 3001 Leuven, Belgium}
\affiliation{Politecnico di Milano, Dipartimento di Chimica, Materiali e Ingegneria Chimica ``Giulio Natta'', 20131 Milano, Italy}

\author{Davide Tierno}
\affiliation{Imec, 3001 Leuven, Belgium}

\author{Kristof Moors}
\affiliation{Forschungszentrum J\"{u}lich, Peter Gr\"{u}nberg Institut, 52428 J\"{u}lich, Germany}

\author{Zsolt T\H{o}kei}
\author{Christoph Adelmann}
\email[Author to whom correspondence should be addressed. Electronic Mail: ]{christoph.adelmann@imec.be}
\affiliation{Imec, 3001 Leuven, Belgium}

\begin{abstract}

The temperature coefficients of the resistivity (TCR) of Cu, Ru, Co, Ir, and W thin films have been investigated as a function of film thickness below 10 nm. Ru, Co, and Ir show bulk-like TCR values that are rather independent of the thickness whereas the TCR of Cu increases strongly with decreasing thickness. Thin W films show negative TCR values, which can be linked to high disorder. The results are qualitatively consistent with a temperature-dependent semiclassical thin-film resistivity model that takes into account phonon, surface, and grain boundary scattering. The results indicate that the thin-film resistivity of Ru, Co, and Ir is dominated by grain boundary scattering whereas that of Cu is strongly influenced by surface scattering.

\end{abstract}

\maketitle 

The perennial scaling of complementary metal-oxide-semiconductor (CMOS) circuits requires the equal miniaturization of the interconnect wires that link individual transistors.\cite{M_2003, CGJ_2014, BAZ_2015} Today, interconnect dimensions have reached about 15 nm and are expected to reduce below 10 nm in the near future. At such small dimensions, the currently used Cu metallization suffers from a strongly increased resistance due to finite size effects of the resistivity \cite{JBT_2009} and scaling limitations of the barriers and liners required to ensure the interconnect reliability.\cite{CWK_2015, O_2015} As a result, the overall performance of CMOS circuits is increasingly limited by the interconnect.\cite{KMS_2002} This has prompted much research to find alternative metals that could replace Cu with both improved reliability and resistivity at small dimensions. Recently, this has led to the introduction of Co in local interconnects.\cite{AAA_2017}

Although Cu has a lower bulk resistivity than the proposed alternative metals, it has been argued\cite{KMS_2002,PN_2014,G_2016} and later experimentally observed \cite{WRP_2016,DSM_2017} that metals with a shorter mean free path of the charge carriers can outperform Cu in thin films or narrow wires due to a reduced sensitivity to surface and grain boundary scattering. However, there is still no consensus on the relative importance of the various scattering contributions and their material dependence. The understanding of the relative importance of scattering mechanisms is crucial for the optimization of the interconnect resistance and thus a simple and robust measurement methodology is desirable. Typically, surface and grain boundary scattering have been modeled as a function of film thickness and grain size within semiclassical approaches\cite{F_1938,S_1952,S_1967,MS_1970} but the disentanglement of the different scattering contributions is not straightforward since the resulting thickness dependences are rather similar and Matthiessen's rule does not apply.\cite{MS_1970,DSM_2017} Temperature-dependent resistivity measurements have been proposed as a possible improvement\cite{AWV_1984,dV_1988,ME_2004} since semiclassical models describe the thin-film resistivity by the ratio of film thickness or grain size to the mean free path of the bulk metal. The mean free path varies with temperature and therefore temperature-dependent resistivity measurements can be used to test the applicability and consistency of semiclassical models and their parameters beyond the description of the simple thickness dependence of the resistivity. Moreover, it has been shown that temperature-dependent resistivity measurements are capable to distinguish qualitatively between dominant surface or grain boundary scattering. Concretely, dominant surface scattering typically leads to a stronger temperature dependence of the resistivity whereas grain boundary scattering leaves it unaffected.\cite{ME_2004} However, only few temperature-dependent thin-film resistivity measurements have been reported on a limited set of materials and the experimental results have not been yet systematically compared to semiclassical models.\cite{KBS_2004,PAD_2006,ZCZ_2006, CLS_2014,CLX_2015} Hence, no consistent picture has emerged yet.
 
Here, we report on the temperature coefficients of the resistivity (TCR) of Cu, Ru, Co, Ir, and W films with thicknesses between 3 and 10 nm. The experimentally measured linear TCR values at room temperature are compared to results of a temperature-dependent semiclassical model for thin-film resistivities. Good qualitative agreement between experiment and model was observed although the magnitude of the observed variation was different for Cu. This demonstrates both the relevance and the quantitative limitations of semiclassical models to describe thin-film resistivities.

\begin{figure*}[tb]
  \includegraphics[width=16.2cm]{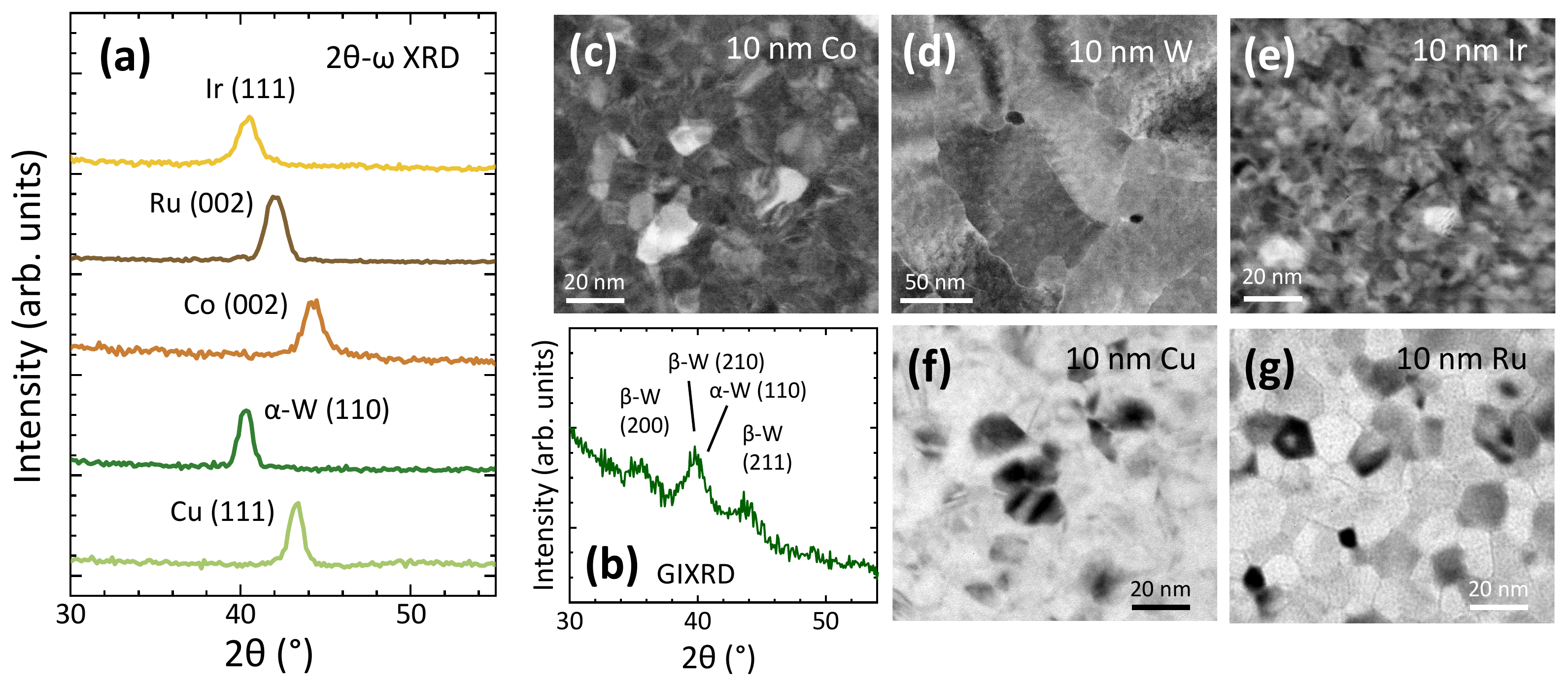}
  \caption{\label{Fig_phys}(a) $2\theta$--$\omega$ x-ray diffraction pattern of the studied 10 nm thick metal films, as indicated. (b) Grazing-incidence x-ray diffraction pattern of 3 nm thick W film. Plan-view dark-field scanning transmission electron micrographs of 10 nm thick (c) Co, (d) W, and (e) Ir films. (f) and (g) show plan-view bright-field transmission electron micrographs of 10 nm thick Cu and Ru films, respectively.}
\end{figure*} 
 
All films were deposited by physical vapor deposition (PVD) in a Canon Anelva EC7800 system at room temperature on 300 mm Si (100) wafers. Prior to metal deposition, a 100 nm thick thermal SiO$_2$ was grown to ensure electrical isolation. Ru, Ir, and W were directly deposited on SiO$_2$, whereas Co and Cu were sandwiched between 1.5 nm thick TaN layers \emph{in situ} to avoid oxidation in air. The parallel conductance of the TaN layers was negligible. Film thicknesses were determined by a combination of x-ray reflectivity and Rutherford backscattering spectrometry. X-ray diffraction (XRD, $2\theta$--$\omega$ geometry, Cu K$\alpha$ radiation, Fig.~\ref{Fig_phys}a) indicated that the films were polycrystalline with strong (111), (110), (001) texture for the fcc (Cu, Ir), bcc (W), and hcp (Co, Ru) metals, respectively. For W, the appearance of the $\beta$-W phase was observed for the thinnest films by grazing-incidence x-ray diffraction (GIXRD, $\omega = 0.3^\circ$, Cu K$\alpha$ radiation, Fig.~\ref{Fig_phys}b). The rms surface roughness measured by atomic force microscopy was 3--5 \AA{} for all films (not shown). Linear intercept lengths between grain boundaries were determined from plan-view transmission electron micrographs (Figs.~\ref{Fig_phys}b--\ref{Fig_phys}f).\cite{DSM_2017} Thin-film resistivities were obtained using both patterned Hall bars and sheet resistance measurements. The TCR was obtained from the Hall bar resistivity at temperatures between 25 \degree C and 125 \degree C. In the studied temperature window, the resistivity was found to increase linearly with temperature within experimental precision, \emph{i.e.}~the TCR was approximately constant. 

\begin{figure}[t]
  \includegraphics[width=8.2cm]{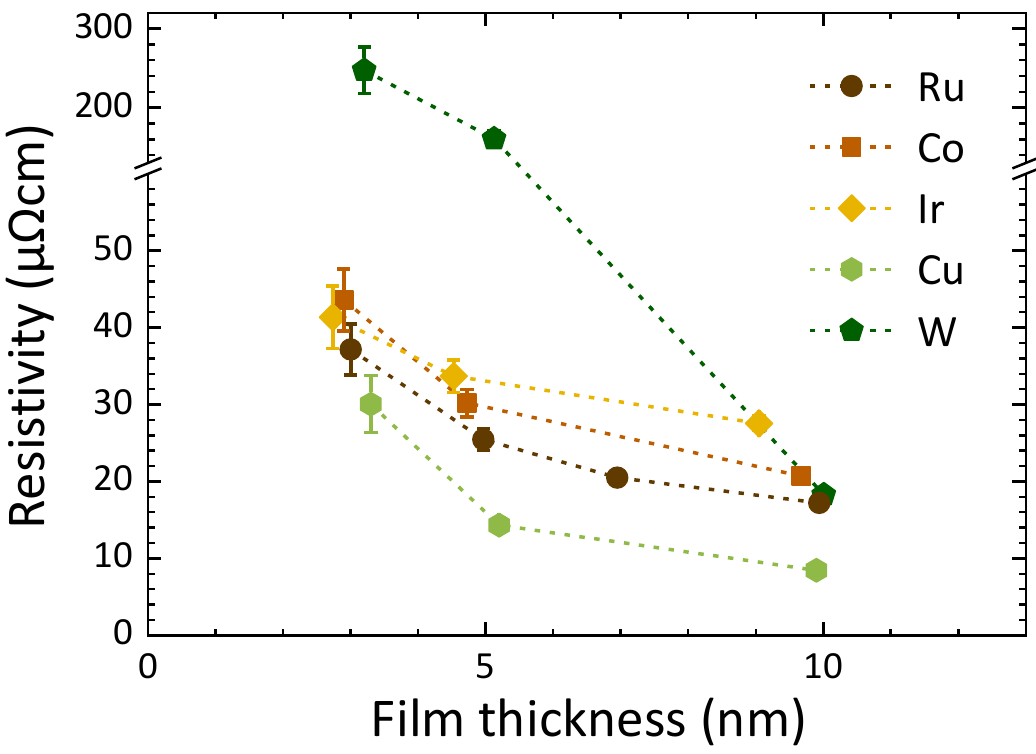}
  \caption{\label{Fig_res}Experimental room-temperature resistivities of the studied thin films as a function of their thickness.}
\end{figure}

\begin{figure}[tb]
  \includegraphics[width=8.3cm]{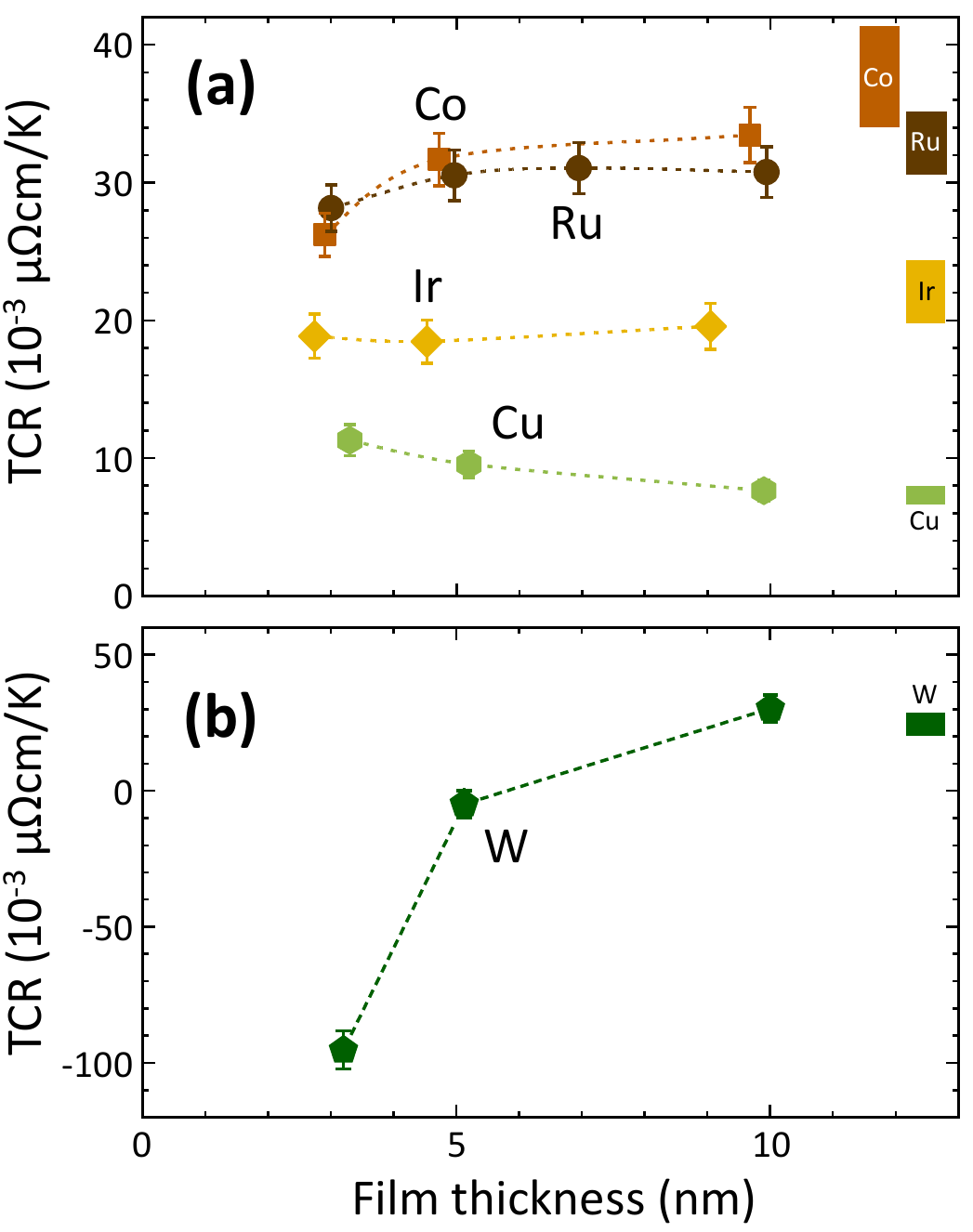}
  \caption{\label{Fig_tcr}Experimental TCR values near room-temperature of the studied thin films as a function of their thickness. The boxes on th right hand side of the graphs indicate the range of bulk TCR values in Ref.~\onlinecite{BHO}.}
\end{figure}

Figure~\ref{Fig_res} shows the measured resistivities of the different metal thin films at room temperature as a function of their thickness. For all cases, the resistivity increased with decreasing thickness due to increasing contributions of surface and grain boundary scattering in thinner films. For thicknesses of 5 nm and above, Cu had clearly the lowest resistivity. However, for 3 nm thick films, resistivities of alternative metals (except W) became comparable, in keeping with previous reports \cite{DSM_2017}. This has been explained by the longer mean free path of Cu with respect to the other metals,\cite{G_2016} which renders Cu much more sensitive to finite size effects. Figure~\ref{Fig_tcr} shows the experimentally determined TCR of the same set of thin films as a function of their thickness. The TCR of Cu was close to the bulk value of $6.6\times 10^{-3}$ \textmu{}$\Omega$cm \cite{BHO} for the thickest film but increased strongly as the film thickness decreased. For 3 nm thick Cu, the TCR was about 70\%{} higher than the bulk value. TCR values for Ru, Ir, and Co films were close to bulk values\cite{BHO} for thicknesses between 10 and 5 nm, with some reduction below the bulk value for the thinnest Ru (by about 10\%{}) and Co (by about 20\%{}) films. For Ir, this decrease was absent and even the thinnest film showed a bulk-like TCR within experimental accuracy. By contrast, the behavior of W was distinctly different (Fig.~\ref{Fig_tcr}b). While the TCR was close to the bulk value for the 10 nm thick film, it decreased sharply with decreasing thickness to a strongly negative value at 3 nm film thickness.

To shed light on the experimental observations, the temperature dependence of the thin-film resistivity was calculated using a semiclassical model based on the work by Mayadas and Shatzkes (MS).\cite{MS_1970,ME_2004,A_2019} In the MS model, the thickness dependence of the resistivity in presence of surface and grain boundary scattering is given by

\begin{equation} \label{eq:MS_model}
    \rho_{MS} = \left\{\frac{1}{{{\rho }_{GB}}}-\frac{6}{\pi {{k}}{{\rho }_{0}}}(1-p)\int_{0}^{\pi /2}{d\varphi \int_{1}^{\infty }{dt}\frac{{{\cos }^{2}}\varphi }{{{H}^{2}\left(\varphi, t\right)}}}\times\left( \frac{1}{{{t}^{3}}}-\frac{1}{{{t}^{5}}}   \right)\frac{1-{{e}^{-{{k}}tH\left(\varphi, t\right)}}}{1-p{{e}^{-{{k}}tH\left(\varphi, t\right)}}} \right\}^{-1}\, ,
\end{equation}

\noindent with the abbreviations $\rho_{GB}=\rho_{0} \left[ 1-3\alpha /2+3\alpha^2 - 3\alpha^3\ln \left( 1 + 1/\alpha \right) \right]^{-1}$, $\alpha = \frac{{{\lambda }_{0}}}{g}\frac{2R}{1-R}$, and $H\left(\varphi, t\right) = 1 + \alpha / \cos\varphi\sqrt{\left( 1 - 1/t^2 \right)}$. Here, $\rho_0$ is the bulk resistivity of the metal, $h$ the film thickness, $\lambda$ the mean free path of the charge carriers, $k={h}/{\lambda}$, and $g$ the mean linear intercept length between grain boundaries. $0 \le R \le 1$ is the grain boundary reflection coefficient and determines the strength of grain boundary scattering. The parameter $p$ describes the scattering at the surfaces or interfaces of the films with a value of 0 corresponding to fully diffuse and 1 to fully specular scattering.

The MS model does not depend explicitly on the temperature $T$, but implicitly via the bulk mean free path $\lambda(T)$ and the bulk resistivity $\rho_0(T)$. It has however been shown that the product $\rho_0\times \lambda \equiv A$ is a function of the Fermi surface morphology only and can be calculated by \emph{ab initio} methods.\cite{G_2016,DSM_2017} Moreover, the product $A$ is independent of temperature for $T\ll T_F$ with $T_F$ the Fermi temperature of the metal. The temperature dependence of the bulk resistivity $\rho_0(T)$ in presence of phonon and (weak) impurity scattering can be described by the Bloch-Gr\"{u}neisen model 

    \begin{equation} \label{eq:bloch-gruneisen}
    {{\rho }_{0}}(T)=\rho_\mathrm{imp}+C{{T}^{5}}\int_{0}^{{{{\Theta }_{D}}}/{T}}{\frac{{{x}^{5}}}{({{e}^{x}}-1)(1-{{e}^{-x}})}dx}\, ,
    \end{equation} 

\noindent where $\rho_\mathrm{imp}$ describes the residual (temperature-independent) resistivity due to impurity or point defect scattering. $\Theta_D$ is the Debye temperature, and $C$ is a prefactor that can be determined from the bulk room-temperature resistivity. In high-purity PVD films, impurity scattering can be neglected at room temperature. The temperature dependence of the mean free path $\lambda(T)$ can then be calculated by $\lambda(T) = A/\rho_0(T)$. This is equivalent to assuming that the carrier density in the metal is independent of temperature and the temperature dependence of the resistivity is determined by scattering only, which is generally well obeyed in metals. Equation~(\ref{eq:bloch-gruneisen}) then allows for the calculation of $\lambda(T)$, which in turn can be used to calculate the temperature-dependent thin-film resistivity by Eq.~(\ref{eq:MS_model}).\cite{A_2019} An analytical model of the TCR based on this approach has been published by Marom and Eizenberg.\cite{ME_2004} However, it is straightforward to calculate the temperature-dependent resistivity numerically using Eq.~(\ref{eq:MS_model}) and to obtain the TCR by differentiation. The materials parameters used for the calculation of the TCR of the different metal films are listed in Tab.~\ref{tab:input}. 

\begin{table}[bt]
\caption{\label{tab:input} Material parameters used for modeling the TCR: room-temperature bulk resistivity $\rho_{0,\mathrm{rt}}$, room-temperature mean free path $\lambda_{0,\mathrm{rt}}$, temperature-independent $\rho_{0}\times\lambda_{0}$ product,\cite{G_2016,DSM_2017} Debye temperature $\Theta_D$,\cite{Debye_book} grain boundary scattering parameter $R$, and surface scattering parameter $p$.\cite{JBT_2009,DSM_2017,CLS_2014,DBG_2018}}
\begin{ruledtabular}
\begin{tabular}{p{10pt}p{28pt}p{20pt}p{45pt}p{25pt}p{12pt}p{12pt}}

\parbox{10pt}{\centering } & 
\parbox{28pt}{\centering $\rho_{0,\mathrm{rt}}$ (\textmu$\Omega{}$cm)} &
\parbox{20pt}{\centering $\lambda_{0,\mathrm{rt}}$ (nm)} & \parbox{45pt}{\centering $\rho_0\times\lambda_0$ ($10^{-16} \Omega$m\textsuperscript{2})} &
\parbox{25pt}{\centering $\Theta_{D}$ (K)} & 
\parbox{12pt}{\centering $R$} & 
\parbox{12pt}{\centering $p$} \\
\hline
\parbox{10pt}{\centering Cu} & 
\parbox{28pt}{\centering 1.7} & 
\parbox{20pt}{\centering 39.9} & 
\parbox{45pt}{\centering 6.70} & 
\parbox{25pt}{\centering 320} & 
\parbox{12pt}{\centering 0.22} & 
\parbox{12pt}{\centering 0} \\

\parbox{10pt}{\centering Co} & 
\parbox{28pt}{\centering 6.2} & 
\parbox{20pt}{\centering 7.8} & 
\parbox{45pt}{\centering 4.82} & 
\parbox{25pt}{\centering 365} & 
\parbox{12pt}{\centering 0.37} & 
\parbox{12pt}{\centering 0} \\

\parbox{10pt}{\centering Ru} & 
\parbox{28pt}{\centering 7.8} & 
\parbox{20pt}{\centering 6.6} & 
\parbox{45pt}{\centering 5.14} & 
\parbox{25pt}{\centering 385} & 
\parbox{12pt}{\centering 0.50} &
\parbox{12pt}{\centering 1} \\

\parbox{10pt}{\centering Ir} & 
\parbox{28pt}{\centering 5.2} & 
\parbox{20pt}{\centering 7.1} & 
\parbox{45pt}{\centering 3.69} & 
\parbox{25pt}{\centering 285} & 
\parbox{12pt}{\centering 0.50} &
\parbox{12pt}{\centering 1} \\

\parbox{10pt}{\centering W} & 
\parbox{28pt}{\centering 5.3} & 
\parbox{20pt}{\centering 15.5} & 
\parbox{45pt}{\centering 8.20} & 
\parbox{25pt}{\centering 320} & 
\parbox{12pt}{\centering 0.55} & 
\parbox{12pt}{\centering 0} \\
\end{tabular}
\end{ruledtabular}
\end{table}

\begin{figure}[tb]
  \includegraphics[width=8.3cm]{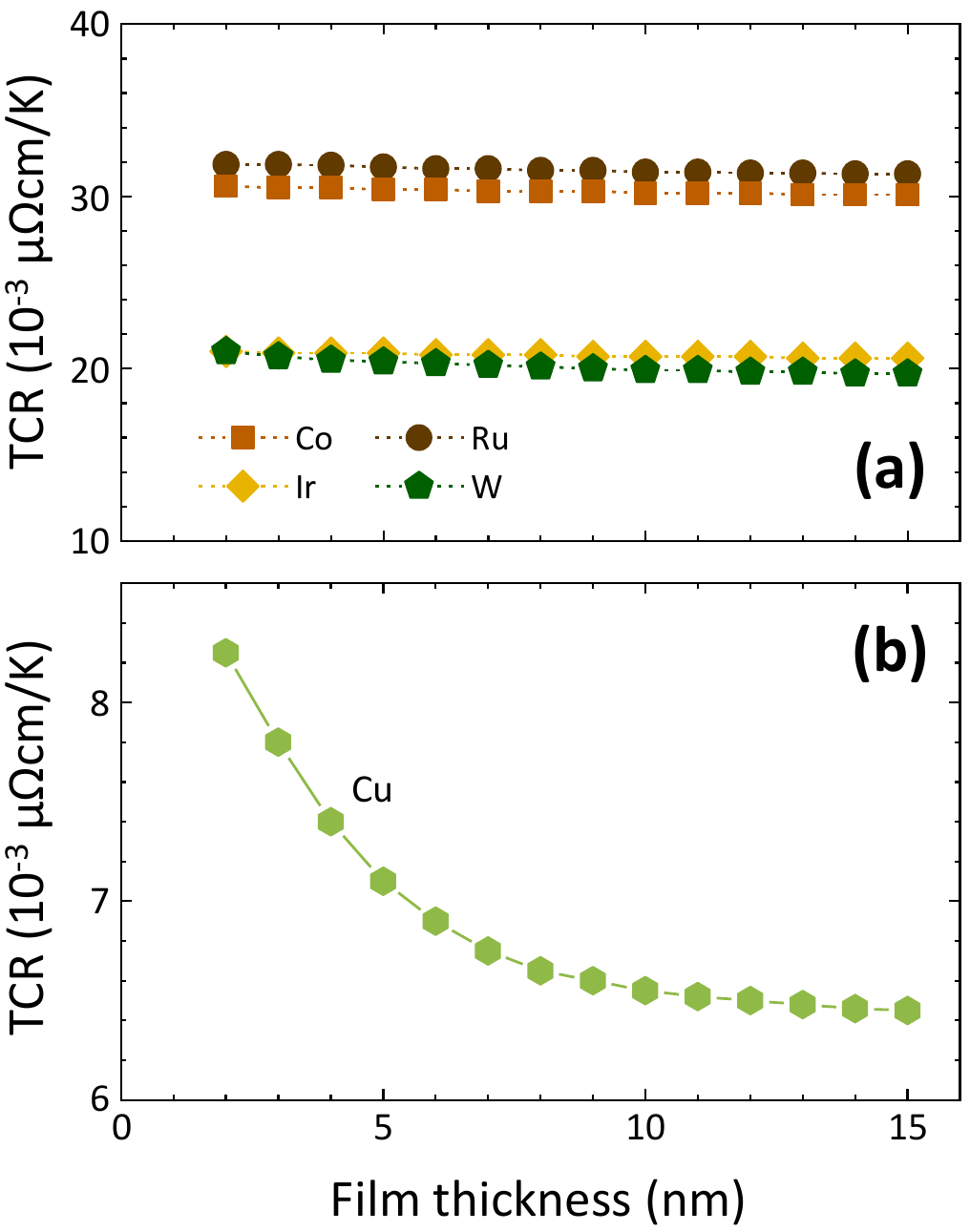}
  \caption{\label{Fig_tcr_calc} Calculated film thickness dependence of the TCR at 300 K for (a) Co, Ir, Ru, and W as well as (b) Cu.}
\end{figure}

In general, the calculated TCR decreased weakly (by about 2\%) between 300 K and 400 K, which is below experimental precision and therefore the TCR at 300 K is reported for simplicity. Values for the different metals are shown as a function of film thickness in Fig.~\ref{Fig_tcr_calc}. For Co, Ru, and Ir, the calculated TCR values were independent of thickness (less than 5\% variation) and within 3\% of the calculated bulk value, in good agreement with the experimental results. This indicates that the increase of the thin-film resistivity with decreasing thickness is independent of temperature. Such a behavior has been linked to cases where the thin-film resistivity is dominated by grain boundary scattering.\cite{A_2019, ME_2004} The results for Ru confirm a previous analysis of the thickness dependence of the Ru thin-film resistivity,\cite{DSM_2017} indicating the dominance of grain boundary scattering. Similar to scattering by point defects, quantum-mechanical tunneling through grain boundaries is expected to depend only very weakly on temperature, which is consistent with both the modeled and experimentally observed behavior.

In addition, the increase of the experimental TCR of a 3 nm thick Cu film by about 70\% over the bulk could also be qualitatively explained within the semiclassical model. In this model, an increasing TCR for decreasing film thickness has been found in cases of strong contributions of surface scattering to the thin-film resistivity.\cite{A_2019, ME_2004} This confirms a previous analysis of the thickness dependence of the Cu thin-film resistivity.\cite{DSM_2017} However, the measured increase of the thin-film TCR with respect to the bulk value was about three times as large as the calculated one. This suggests that the above semiclassical model only describes qualitative aspects of the resistivity of thin metallic films in presence of surface scattering. Such limitations may stem from various sources, such as the assumption of an isotropic free electron gas or from the omission of point defect scattering and quantum confinement effects in the semiclassical model. These results may also qualitatively explain a previous report, which found that the thickness dependence of the Cu thin-film resistivity required different fitting parameters $p$ and $R$ at different temperatures.\cite{PAD_2006} While it cannot be ruled out that $p$ and $R$ depend indeed on temperature, our findings suggest that the discrepancies may at least partially stem from limitations of the model to accurately and consistently describe thin-film resistivities at different temperatures. The results also suggest that temperature-dependent measurements are well suited to test the accuracy of future improved models of thin-film or nanowire resistivity.

By contrast, the negative TCR for W (Fig.~\ref{Fig_tcr}b) cannot be explained within the semiclassical model for metallic thin films described above. The semiclassical model predicts that the W thin-film resistivity increases weakly with decreasing thickness, which stems from a nonnegligible influence of surface scattering due to the relatively long mean free path of W (Fig.~\ref{Fig_tcr_calc}a). Lower-than-bulk and even negative TCR values have however been observed for highly resistive metals, especially with resistivities around or above the Ioffe-Regel limit.\cite{M_1973, I_1980, T_1986} The behavior has been attributed to localization effects due to large disorder and a breakdown of Matthiessen's rule between point defect or grain boundary scattering and phonon scattering.\cite{M_1973, I_1980, T_1986, CDD_2018} Charge carrier localization leads generally to a weaker temperature dependence of the resistivity. Thermal activation effects when localization energies become comparable to the thermal energy can even lead to negative TCR values. Moreover, in the case of strong disorder, contributions of impurity or grain boundary scattering and phonon scattering are not additive anymore and cannot be clearly separated. Since both the semiclassical MS model in Eq.~(\ref{eq:MS_model}) and the Bloch-Gr\"{u}neisen model in Eq.~(\ref{eq:bloch-gruneisen}) explicitly assumes the validity of Matthiessen's rule between phonon and grain boundary or impurity scattering,\cite{MS_1970} such effects cannot be described within the above approach. 

For W, the large disorder for the thinnest films may be linked to the appearance of the high-resistivity $\beta$-W phase, as demonstrated in Fig.~\ref{Fig_phys}b. The formation of $\beta$-W has been typically observed for PVD films below a certain critical thickness, typically between 5 and 20 nm,\cite{CWC_2011,HCX_2015} depending on the deposition conditions. We note that such negative TCR values were not observed for W deposited by chemical vapor deposition.\cite{MKZ_2018} 

The same disorder and localization effects leading to the breakdown of Matthiessen's rule between point defect or grain boundary scattering and phonon scattering may also explain the observed reduction of the TCR of the thinnest Ru and Co films. Films deposited by PVD often contain a disordered nanocrystalline interface layer at the substrate due to random nucleation, limited adatom mobility, and/or high stress. The disorder in such ultrathin nanocrystalline may be due to point defects but also due to a high density of grain boundaries. All these effects can lead to (weak) localization of the charge carriers close to the interface and the observed reduction of the TCR.

In conclusion, we have studied the TCR of Cu, Co, Ru, Ir, and W thin films with thicknesses between 3 and 10 nm. The TCR of Co, Ru, and Ir was bulk-like except for the thinnest films, where the TCR was slightly reduced. By contrast, the TCR of Cu increased with decreasing thickness and became larger than the bulk value. These observations could be qualitatively explained by a semiclassical model for the temperature dependence of the thin-film resistivity. In agreement with a previous analysis of the thickness dependence of the thin-film resistivity,\cite{DSM_2017} the model was consistent with the predominance of grain boundary scattering in Co, Ru, and Ir, whereas the behavior of Cu was influenced by a strong contribution of surface scattering. By contrast, the TCR of W became strongly negative for the thinnest films, indicating the presence of strong disorder, presumably due to the appearance of the high-resistivity $\beta$-W phase. 

The results indicate that semiclasssical thin-film resistivity models\cite{MS_1970} can describe the TCR qualitatively without the need of assuming temperature-dependent model parameters. However, the models fail to describe the measured thickness- and temperature-dependence quantitatively in a consistent way for predominant surface scattering. This hints towards limitations of such semiclassical models to describe the resistivity of thin metallic films in all cases fully quantitatively. Improved models, \emph{e.g.}~taking the band structure into account, may thus be required for a quantitative consistent picture of the thin-film resistivity and its thickness and temperature dependence.

This work has been supported by imec's industrial affiliation program on nano-interconnects. The authors would like to thank Sofie Mertens and Thomas Witters (imec) for the support of the PVD depositions as well as imec's Materials and Components Analysis (MCA) Laboratory for the electron micrographs, the atomic-force microscopy, and the Rutherford backscattering measurements. All data are available on request from the authors.

\end{document}